\documentclass[12pt,graphicx]{article}

\usepackage{amssymb}

\usepackage{amsmath}

\newtheorem{theorem}{Theorem}

\newcommand{\der}[2]{\frac{\partial #1}{\partial #2}}

\newcommand{\dertot}[2]{\frac{d #1}{d #2}}

\newcommand{\paper}[6]{#1, #2, {\em #3}, {\bf #4}, #5,  (#6)}

\newcommand{\book}[4]{#1, {\em #2}, {#3}, (#4)}

\begin{document}
 
\title{Symmetry constraints for real dispersionless
Veselov-Novikov equation}

\date{}

\author{Leonid Bogdanov\footnote{Landau Institute for Theoretical
Physics, RAS, Moscow, Russia, e-mail: leonid@landau.ac.ru}, Boris
G. Konopelchenko\footnote{Dipartimento di Fisica dell'
Universit\`{a} di Lecce and INFN, Sezione di Lecce, I-73100 Lecce,
Italy, e-mail: konopel@le.infn.it} ~ and Antonio
Moro\footnote{Dipartimento di Fisica dell' Universit\`{a} di Lecce
and INFN, Sezione di Lecce, I-73100 Lecce, Italy, e-mail:
antonio.moro@le.infn.it}}

\maketitle

\begin{abstract}
Symmetry constraints for dispersionless integrable equations are
discussed. It is shown that under symmetry constraints the
dispersionless Veselov-Novikov equation is reduced to the
$1+1$-dimensional hydrodynamic type systems. \\

\end{abstract}

\section{Introduction.}
Dispersionless integrable equations have attracted recently a
considerable interest (see e.g.~\cite{Zakharov}-\cite{Singular}). They arise in various fields
of physics, mathematical physics and applied mathematics.
Several methods and approaches have been used to
study dispersionless systems, from the
quasi-classical Lax pair representation with its close relationship with
the Whitham universal hierarchy~\cite{Krichever1,Krichever2} to the
quasi-classical version of the inverse scattering method. In
particular, the quasi-classical
$\bar{\partial}$-dressing method~\cite{Ragnisco,Konopelchenko1,Konopelchenko2}, recently formulated,
is a general and
systematic approach to construct dispersionless integrable systems and
to find their solutions. On the other hand a reduction method (see
e.g. \cite{Kodama,KodGibb,Gibbons}) provides us also with the effective way to solve
$2+1$-dimensional dispersionless equations.
It was shown~\cite{Bogdanov} that symmetry constraints for
dispersionless equations
provide us with an efficient way to construct
reductions.
In~\cite{Bogdanov} certain hydrodynamic type reductions of dispersionless
Kadomtsev-Petviashvili (dKP) and dispersionless two-dimensional Toda
Lattice (2DdTL) equations
have been constructed using symmetry constraints.

In this paper we study\quad{}symmetry constraints\quad{}for dispersionless
Veselov-Novikov equation (dVN) and analyze the corresponding
hydrodynamic type equations.

In section~\ref{sec_symmetry} we remind the definition of {\em symmetry
constraint}. In 
sections~\ref{sec_soliton}~and~\ref{sec_displess} symmetry
constraints for soliton (dispersive) and dispersionless equations
respectively are considered,
where the Kadomtsev-Petviashvili (KP) and dKP equation are considered
as examples. 
Symmetry constraints for real dVN
equation are discussed in section~\ref{sec_dVN}. In section~\ref{sec_hydro} we
demonstrate how symmetry constraints for dVN equation allow us to reduce this $2+1$-dimensional
equation to $1+1$-dimensional hydrodynamic type systems.

\section{Symmetry constraints.}
\label{sec_symmetry}
Let us consider a partial differential equation for the
scalar function $u = u(t) = u \left(t_{1},t_{2},\dots \right)$
\begin{equation}
\label{general} F\left(u,u_{t_{i}},u_{t_{i}t_{j}},\dots
\right) = 0,
\end{equation}
where $u_{t_{i}} = \partial u/\partial t_{i}$. By definition, a symmetry of
equation~(\ref{general}) is a transformation $u(t) \rightarrow
u'(t')$, such that $u'(t')$ is again a solution of~(\ref{general}) (for
more details see e.g.~\cite{ClassSymEng}).
Infinitesimal continuous symmetry transformations
\begin{equation}
t_{i}'=t_{i}+\delta t_{i};~~~~~u'=u + \delta u = u + \epsilon
u_{\epsilon}.
\end{equation}
are defined by the linearized equation~(\ref{general})~\cite{ClassSymEng}
\begin{equation}
\label{symmetry_def} L\delta u = 0,
\end{equation}
where $L$ is the
Gateaux derivative of $F$
\begin{equation}
\label{gateaux} L\delta u := \left . \dertot{F}{\epsilon} \left(u + \epsilon
u_{\epsilon},\der{}{t_{i}}\left(u+\epsilon
u_{\epsilon}\right),\dots),\dots \right) \right |_{\epsilon = 0}.
\end{equation}
Any linear
superposition $\delta u =\sum_{k} c_{k} \delta_{k}
u$ of infinitesimal symmetries $\delta_{k}u$ is, obviously, an
infinitesimal symmetry. By definition a {\em symmetry constraint} is
a requirement that certain superposition of infinitesimal symmetries
vanishes, i.e.
\begin{equation}
\label{constraint_gen} \sum_{k} c_{k} \delta_{k} u = 0.
\end{equation}
Since null function is a symmetry of
equation~(\ref{general}), the constraint~(\ref{constraint_gen}) is
compatible with equation~(\ref{general}). Symmetry constraints allow us
to select a class
of solutions which possess some invariance properties. For
instance, well-known symmetry constraint $\delta u = \epsilon u_{t_{k}} = 0$,
selects solutions which are stationary with respect to the ``time'' $t_{k}$.

\section{Soliton equations.}
\label{sec_soliton}
Symmetry constraints for $2+1$-dimensional soliton equations have been discussed the
first time in the papers~\cite{Strampp,Cheng}. Here, we discuss the KP equation
equation~(see e.g. \cite{Segur})
\begin{align}
\label{KP_full}
u_{t} &= \frac{3}{2}u u_{x} + u_{xxx} +
\frac{3}{4}
\omega_{y} \nonumber\\
\omega_{x} &= u_{y},
\end{align}
where $x:=t_{1}$, $y:=t_{2}$ and $t:=t_{3}$. KP
equation~(\ref{KP_full}) is equivalent to the compatibility of the
following linear problems~\cite{Segur}
\begin{align}
\label{KP_lax}
\psi_{y}& = \psi_{xx} + u \psi \nonumber \\
\psi_{t}& = \psi_{xxx} + \frac{3}{2} u \psi_{x} + \left(
\frac{3}{2} u_{x} + \frac{3}{4} \omega \right) \psi.
\end{align}
The symmetries equation~(\ref{symmetry_def}) for KP assumes the
form
\begin{align}
\label{symmetry_KP} 
(\delta u)_{t}& = \frac{3}{2} \left(u_{x}
\delta u + u (\delta u)_{x}\right) + (\delta u)_{xxx} +
\frac{3}{4} (\delta
\omega)_{y}, \\
(\delta \omega)_{x}& = (\delta u)_{y}.
\end{align}
Now, introducing the adjoint linear problems of~(\ref{KP_lax})
defined as
\begin{align}
\label{KP_lax_adj}
-\psi^{\ast}_{y}& = \psi^{\ast}_{xx} + u \psi^{\ast} \nonumber\\
\psi^{\ast}_{t}& = \psi^{\ast}_{xxx} + \frac{3}{2} u
\psi^{\ast}_{x} + \left( \frac{3}{2} u_{x} - \frac{3}{4} \omega
\right) \psi^{\ast},
\end{align}
one can verify directly that the function $\phi = \left(\psi
\psi^{\ast}\right)_{x}$ obeys the linearized KP
equation~(\ref{symmetry_KP})~\cite{Orlov}, i.e. $\left(\psi \psi^{\ast}
\right)_{x}$ is an infinitesimal symmetry of the KP equation. A class of symmetry
constraint can be taken as
\begin{equation}
u_{t_{n}} = \left(\psi \psi^{\ast} \right)_{x},~~~~~n=1,2,3.
\end{equation}
The simplest of them is
\begin{equation}
\label{simple_constraint_full} u_{x} = \left(\psi \psi^{\ast}
\right)_{x},
\end{equation}
which can be integrated to
\begin{equation}
\label{integrated}
 u = \psi \psi^{\ast}.
\end{equation}
Substituting~(\ref{integrated}) in the first equation of~(\ref{KP_lax}) and
its adjoint~(\ref{KP_lax_adj}), one obtains
\begin{align}
\label{AKNS1}
\psi_{y} + \psi_{xx} + \psi^{2}\psi^{\ast}& = 0 \\
-\psi^{\ast}_{y} + \psi^{\ast}_{xx} + (\psi^{\ast})^{2}\psi& = 0,
\end{align}
that is the AKNS\quad{}system~\cite{Segur}, which is reduced to the
nonlinear Schr\"odinger equation if $\psi^{\ast} = \bar{\psi}$,
where the ``bar" means complex conjugation.

Substituting~(\ref{integrated}) in the second equations of linear problems
and its adjoint and observing that $\omega = \psi^{\ast}_{x}\psi -
\psi_{x}\psi^{\ast}$, one gets the higher AKNS system
\begin{align}
\label{AKNS2}
\psi_{t} &= 3 \psi \psi^{\ast} \psi_{x} + \psi_{xxx} \\
\label{AKNS2bis}
\psi^{\ast}_{t} &= 3 \psi \psi^{\ast} \psi^{\ast}_{x} +
\psi^{\ast}_{xxx}.
\end{align}
It is a straightforward check that if $\psi$ and $\psi^{\ast}$ obey
equations~(\ref{AKNS1})-(\ref{AKNS2bis}), then $u=\psi \psi^{\ast}$
solves KP equation.
Thus, symmetry constraints can be used to find solutions of
$2+1$-dimensional system using solutions of the $1+1$-dimensional
integrable systems.

\section{Nonlinear dispersionless equations.}
\label{sec_displess}
The dispersionless limit of soliton equations can be performed
introducing slow variables, formally substituting $t_{n} \to
\epsilon^{-1}t_{n}$, and looking for solutions having a
certain behavior when $\epsilon \to 0$, for instance
\begin{equation}
u\left(\frac{t_{n}}{\epsilon} \right) \to u(t_{n}) + O(\epsilon),~~~~~\epsilon \to 0.
\end{equation}
For example, the dispersionless limit of KP equation is
\begin{align}
\label{dKP} 
u_{t}& = \frac{3}{2}u u_{x} + \frac{3}{4}
\omega_{y} \nonumber\\
\omega_{x}& = u_{y}.
\end{align}
The dispersionless limit of an integrable equation corresponds to
the quasiclassical limit of the corresponding linear problems. In
fact, representing 
the solution $\psi$ of~(\ref{KP_lax}) as
\begin{equation}
\psi = \psi_{0} \exp \left( \frac{S}{\epsilon} \right),
\end{equation}
where
$S\left(\lambda;\frac{x}{\epsilon},\frac{y}{\epsilon},\frac{t}{\epsilon}
\right) \to S\left(\lambda;x,y,t \right) + O(\epsilon)$,
and $\lambda$ is the so-called {\em spectral parameter}, in the limit
$\epsilon \to 0$ one gets from~(\ref{KP_lax}) the following pair of Hamilton-Jacobi
type equations
\begin{align}
\label{dKP_lax}
S_{y}& = S_{x}^{2} + u  \nonumber\\
S_{t}& = S_{x}^{3} + \frac{3}{2} u S_{x} +\frac{3}{2} u_{x}+
\frac{3}{4}\omega,
\end{align}
where $\omega_{x} = u_{y}$. The compatibility condition
for~(\ref{dKP_lax}) is just
the dKP equation~(\ref{dKP}). Similarly to the dispersionfull
case, we have the linearized dKP
\begin{align}
 \label{symmetry_dKP}
(\delta u)_{t}& = \frac{3}{2} \left(u_{x}
\delta u + u (\delta u)_{x}\right) + \frac{3}{4} (\delta
\omega)_{y}, \nonumber\\
(\delta \omega)_{x}& = (\delta u)_{y},
\end{align}
whose solutions are infinitesimal symmetries of dKP.
{\theorem Given any solutions $S_{i}$ and $\tilde{S}_{i}$ of the
  Hamilton-Jacobi equations~(\ref{dKP_lax}), the quantity $\delta u =
  \sum_{i=1}^{N} c_{i}\left(S_{i}-\tilde{S}_{i}\right)_{xx} $, where $c_{i}$
are arbitrary constants, is a symmetry of dKP equation.} \\
{\em Proof.} It is straightforward to check that
$\left(S_{i}-\tilde{S}_{i}\right)_{xx}$ satisfies
equation~(\ref{symmetry_dKP}).
{\rightline{$\Box$}} 
This type of symmetries has been introduced for the first time
in~\cite{Bogdanov}, within the quasiclassical
$\bar{\partial}$-dressing approach. 
A simple example of symmetry constraint for dKP, parallel
to~(\ref{simple_constraint_full}), is
\begin{equation}
u_{x} = S_{xx}.
\end{equation}
Under this constraint the Hamilton-Jacobi system~(\ref{dKP_lax})
gives rise~\cite{Bogdanov} to the following hydrodynamic type
system (the dispersionless nonlinear Schr\"odinger equation)~\cite{Zakharov}
\begin{align}
\tilde{u}_{y}& = \left(\tilde{u}^{2} + u \right)_{x},  \nonumber\\
u_{y}& = 2 \left(\tilde{u} u \right)_{x},
\end{align}
where $\tilde{u} = \partial S_{x}/\partial \lambda$.

\section{Real dVN equation.}
\label{sec_dVN}
The Veselov-Novikov (VN) equation has been introduced as the two
dimensional integrable extension of KdV in 1984~\cite{Veselov}.
It looks like
\begin{align}
\label{VN_full}
u_{t} &= \left(u V \right)_{z} + \left(u \bar{V}
\right)_{\bar{z}}
+ u_{zzz} + u_{\bar{z}\bar{z}\bar{z}} \\
V_{\bar{z}}& = -3 u_{z},
\end{align}
where $z = x + i y$. It is equivalent to the compatibility condition for
equations
\begin{align}
&\psi_{z\bar{z}} = u \psi \\
&\psi_{t} = \psi_{zzz} + \psi_{\bar{z}\bar{z}\bar{z}} +
\left(V\psi_{z} \right) + \left(\bar{V}\psi_{\bar{z}} \right).
\end{align}
The VN equation has applications in differential
geometry~\cite{Pinkall,Ferapontov}. Recently, it was shown that the
dVN equation governs the 
propagation of light in a special class of nonlinear media in the
limit of geometrical optics~\cite{KonopMoro}.

The dVN equation can be obtained as slow variables expansion of the
VN equation~(\ref{VN_full}).
Setting $\psi$=$
\psi_{0}$$(\lambda$,$\epsilon^{-1}z$,$\epsilon^{-1}\bar{z}$,$\epsilon^{-1}t)$$\exp \epsilon^{-1}S(\lambda$,$z$,$\bar{z}$,$t)$ just like in the
previous section, one has the following pair of Hamilton-Jacobi
equations~\cite{Krichever1,Konopelchenko2}
\begin{align}
\label{dVN_Lax}
&S_{z}S_{\bar{z}} = u, \\
\label{dVN_Lax_time} &S_{t} = S_{z}^{3}+S_{\bar{z}}^{3} + V S_{z}
+ \bar{V} S_{\bar{z}},
\end{align}
and the equation
\begin{align}
\label{dVN}
u_{t}& = \left(u V \right)_{z} + \left(u \bar{V}
\right)_{\bar{z}}\nonumber \\
V_{\bar{z}}& = -3 u_{z}.
\end{align}
In his paper we consider the case of real-valued $u$. \\
Linearized version of~(\ref{dVN}) is of the form
\begin{align}
\label{linearized_dVN}
\left(\delta u \right)_{t}& = \left(V \delta u + u \delta V
\right)_{z} + \left(\bar{V} \delta u + \delta \bar{V} u
\right)_{\bar{z}} \nonumber \\
V_{\bar{z}}& = -3 u_{z}; \quad{} \left(\delta V \right)_{\bar{z}} =
-3 \left(\delta u \right)_{z}.
\end{align}
{\theorem Given any solutions $S_{i}$ and $\tilde{S}_{i}$ of the
  Hamilton-Jacobi equations~(\ref{dVN_Lax})-(\ref{dVN_Lax_time}), the
  quantity
\begin{equation}
\label{constraint_S}
\delta u =
  \sum_{i=1}^{N} c_{i}\left(S_{i}-\tilde{S}_{i}\right)_{z\bar{z}},
\end{equation}
where $c_{i}$
are arbitrary constants, is a symmetry of dVN equation.} \\
{\em Proof.} It is straightforward to check that
$\left(S_{i}-\tilde{S}_{i}\right)_{z\bar{z}}$ satisfies
equation~(\ref{linearized_dVN}). \\
\rightline{$\Box$} 
In particular, one can
choose $S_{i} = S(\lambda = \lambda_{i})$ and $\tilde{S}_{i} =
S(\lambda = \lambda_{i} + \mu_{i})$. In the case $\mu_{i} \to 0$
and $c_{i} = \tilde{c}_{i}/\mu_{i}$, one has the class of
symmetries given by
\begin{align}
\label{constraint_Phi}
\delta u& = \sum_{i=1}^{N} \tilde{c}_{i} \phi_{iz\bar{z}} \\
\phi_{i}& = \der{S}{\lambda}(\lambda = \lambda_{i}).
\end{align}
In what follows we will discuss three particular cases of real
reductions, providing real solutions of dVN.

If $S$ is a solution of Hamilton-Jacobi
equations~(\ref{dVN_Lax}), then $-\bar{S}$ is a solution as well. Thus,
for real-valued $S$ ($S= \bar{S}$), specializing
constraint~(\ref{constraint_S}) for
$N=1$, we have a simple constraint
\begin{align}
\label{case1} &\text{Case~I} & u_{x} &= \left(S
\right)_{z\bar{z}}.\\
\intertext{For complex valued $S$ one has the constraint}
\label{case2} &\text{Case~II} & u_{x} &= \frac{1}{2}\left(S + \bar{S}
\right)_{z\bar{z}}.\\
\intertext{The last example of constraint is nothing but a particular
case of~(\ref{constraint_Phi}), \textit{i.e.}}
\label{case3} &\text{Case~III} & u_{x} &=
\phi_{z\bar{z}}.
\end{align}

\section{Hydrodynamic type reductions of the dVN equation.}
\label{sec_hydro}
\subsection{Case I}
Let us introduce the functions $\rho_{1}:=S_{x}$ and
$\rho_{2}:=S_{y}$. Thus, the symmetry constraint~(\ref{case1}) can be
written as follows
\begin{equation}
\label{case_I} u_{x} =\frac{1}{4} \left(S_{xx}+S_{yy} \right) =
\frac{1}{4}\left( \rho_{1x} + \rho_{2y} \right).
\end{equation}
In order to analyze
constraint~(\ref{case_I}) it is more convenient to consider
equations~(\ref{dVN_Lax}) in Cartesian coordinates $(x,y)$, i.e.
\begin{align}
\label{dVN_Lax_cartesian1}
&S_{x}^{2} + S_{y}^{2} = 4 u \\
\label{dVN_Lax_cartesian2}
&S_{t} = \frac{1}{4} S_{x}^{3}
-\frac{3}{4} S_{x}S_{y}^{2} + V_{1} S_{x} + V_{2} S_{y},
\end{align}
where $V = V_{1} + i V_{2}$, while dVN equation acquires the form
\begin{align}
\label{dVN_cartesian1}
&u_{t} = \left(u V_{1} \right)_{x} +
\left(u V_{2} \right)_{y}\\
\label{dVN_cartesian2}
&V_{1x} - V_{2y} = - 3 u_{x} \\
\label{dVN_cartesian3} &V_{2x} + V_{1y} = 3 u_{y}.
\end{align}
Substituting~(\ref{dVN_Lax_cartesian1})
into~(\ref{case_I}), one gets the following
hydrodynamic type system
\begin{equation}
\label{hydro_system1} \left( \begin{array}{cc}
\rho_{1}  \\ \rho_{2} \\
\end{array} \right)_{y} =
\left( \begin{array}{cccc}
0 & 1 \\
2 \rho_{1} - 1 & 2 \rho_{2}\\
\end{array} \right)
\left( \begin{array}{cc}
\rho_{1}  \\ \rho_{2} \\
\end{array} \right)_{x}.
\end{equation}

Now, let us focus on definition $V_{\bar z} := - 3
u_{z}$. Differentiating it with respect to $x$, using
constraint~(\ref{case1}) and equations~(\ref{hydro_system1}), one
obtains the equations
\begin{align}
V_{1x}& = - \frac{3}{2} \rho_{1x} + \frac{3}{4} \left( \rho_{1}^{2} +
  \rho_{2}^{2}\right)_{x}  \\
V_{2x}& = \frac{3}{2} \rho_{2x},
\end{align}
which can be trivially integrated providing the following explicit
formulas for $V_{1}$ and $V_{2}$ in terms of $\rho_{1}$ and $\rho_{2}$:
\begin{align}
\label{case1_V}
V_{1}& = - \frac{3}{2} \rho_{1} + \frac{3}{4} \left( \rho_{1}^{2} +
  \rho_{2}^{2}\right) \nonumber  \\
V_{2}& = \frac{3}{2} \rho_{2}.
\end{align}
At this point we can derive $t$-dependent equations for $\rho_{1}$ and
 $\rho_{2}$.
Differentiating equation~(\ref{dVN_Lax_cartesian2}) and using
~(\ref{hydro_system1})~and~(\ref{case1_V}), one obtains the system
\begin{equation}
\left( \begin{array}{cc} \rho_{1}  \\ \rho_{2} \\
\end{array} \right)_{t} =\left( \begin{array}{cc}
A_{11} & A_{12}  \\ A_{21} & A_{22} \\
\end{array} \right)\left( \begin{array}{cc} \rho_{1}  \\ \rho_{2} \\
\end{array} \right)_{x},
\end{equation}
where
\begin{eqnarray*}
&A_{11} = 3 \rho_{1} \left( \rho_{1} -1 \right) ,  &A_{12} = 3 \rho_{2}, \\
&A_{21} = 3 \rho_{2} \left (2 \rho_{1}-1 \right),  &A_{22} = 3
  \rho_{1} \left(\rho_{1}-1 \right) + 6 \rho_{2}^{2}.
\end{eqnarray*}

\subsection{Case II}
In this case (presenting the complex-valued function $S$ in terms of its
real and imaginary parts, $S= \rho + i \varphi$) the symmetry
constraint~(\ref{case2}) acquires the form
\begin{equation}
\label{case_II} u_{x} = \frac{1}{4} \left(\rho_{xx} + \rho_{yy}\right).
\end{equation}
Equation~(\ref{dVN_Lax_cartesian1}) is equivalent to the system
\begin{eqnarray}
\label{case2_lax1}
&&(\nabla \rho)^{2} - (\nabla \varphi)^{2} = 4 u \\
\label{case2_lax2} &&\nabla \rho \cdot \nabla \varphi = 0,
\end{eqnarray}
where $\nabla = (\partial/\partial x,\partial/\partial y)$ and notation $\nabla \rho :=
(\rho_{1},\rho_{2})$ and $\nabla \varphi :=
(\varphi_{1},\varphi_{2})$ is introduced. Let us note that
equation~(\ref{case2_lax2}) allows to express, for instance, the
component $\varphi_{2}$ in terms of the other ones ($\varphi_{2} = -
\rho_{1} \varphi_{1} /\rho_{2}$), so that only the functions $\rho_{2}, \rho_{2}$ and
$\varphi_{1}$ are independent. By using the
constraint~(\ref{case_II}), similar to the previous case, one
shows that $\rho_{1},\rho_{2}$, and $\varphi_{1}$,
satisfy the hydrodynamic system
\begin{equation}
\label{hydro_system2}
\left( \begin{array}{ccc} \rho_{1}  \\ \rho_{2} \\ \varphi_{1} \\
\end{array} \right)_{y} =\left( \begin{array}{ccc}
0 & 1 & 0 \\ a_{1} & a_{2} & a_{3} \\
b_{1} & b_{2} & b_{3} \\
\end{array} \right)\left( \begin{array}{ccc} \rho_{1}  \\ \rho_{2}\\
\varphi_{1} \\
\end{array} \right)_{x},
\end{equation}
where
\begin{align*}
a_{1}& = 2 \rho_{1} \left(1 -
\frac{\varphi_{1}^{2}}{\rho_{2}^{2}} \right) -1, \quad{}a_{2} = 2\left(
\rho_{2} +  \frac{\rho_{1}^{2}
\varphi_{1}^{2}}{\rho_{2}^{3}} \right), \\
a_{3}& = -2 \varphi_{1}
\left(1+ \frac{\rho_{1}^{2}}{\rho_{2}^{2}} \right),\quad{}
b_{1} = - \frac{\varphi_{1}}{\rho_{2}}, \quad{}b_{2} =
\frac{\rho_{1}}{\rho_{2}^{2}} \varphi_{1}, \quad{}b_{3}
=-\frac{\rho_{1}}{\rho_{2}}.  
\end{align*}
Just like in the previous section, starting with the definition of $V$
and differentiating it with respect to $x$,
it possible to express its real and imaginary parts in terms of
$\rho_{1}$, $\rho_{2}$, $\varphi_{1}$ and $\varphi_{2}$
\begin{align}
\label{case2_V}
V_{1}& = - \frac{3}{2} \rho_{1} + \frac{3}{4} \left( \rho_{1}^{2} +
  \rho_{2}^{2} - \varphi_{1}^{2} - \varphi_{2}^{2}\right) \nonumber  \\
V_{2}& = \frac{3}{2} \rho_{2}.
\end{align}
or
\begin{align*}
%\label{case2_Vbis}
V_{1}& = - \frac{3}{2} \rho_{1} + \frac{3}{4} \left( \rho_{1}^{2} +
  \rho_{2}^{2} - \varphi_{1}^{2} - \frac{\rho_{1}^{2}
  \varphi_{1}^{2}}{\rho_{2}^{2}}\right) \\ 
V_{2}& = \frac{3}{2} \rho_{2}.
\end{align*}
Separating real and imaginary parts in
equation~(\ref{dVN_Lax_cartesian2}),
one gets the system
\begin{align}
\rho_{t} &= \frac{1}{4} \left(\rho_{x}^{3} - 3
\rho_{x}\varphi_{x}^{2} \right) - \frac{3}{4} \left
(\rho_{x}\rho_{y}^{2} - \rho_{x} \varphi_{y}^{2} - 2 \rho_{y}
\varphi_{x} \varphi_{y} \right) + V_{1} \rho_{x} + V_{2}
\rho_{y},  \label{case2_lax3} \\
\varphi_{t} &= \frac{1}{4}
\left(-\varphi_{x}^{3} + 3 \rho_{x}^{2} \varphi_{x} \right)-
\frac{3}{4} \left(2 \rho_{x} \rho_{y}\varphi_{y} + \varphi_{x}
\rho_{y}^{2} - \varphi_{x} \varphi_{y}^{2} \right) + V_{1}
\varphi_{x} + V_{2} \varphi_{y}. \label{case2_lax4} 
\end{align}
Substituting expressions~(\ref{case2_V}) into~(\ref{case2_lax3})
and~(\ref{case2_lax4}) and differentiating with respect to $x$ and $y$,
one obtains the  hydrodynamic type system for $\rho_{1}$, $\rho_{2}$
and $\varphi_{1}$ 
\begin{eqnarray}
\label{hydro_system3}
\left( \begin{array}{ccc} \rho_{1}\\ \rho_{2} \\ \varphi_{1}    \\
\end{array} \right)_{t} =\left( \begin{array}{ccc}
B_{11} & B_{12} & B_{13} \\ B_{21} & B_{22} & B_{23} \\
B_{31} & B_{32} & B_{33} \\
\end{array} \right)\left( \begin{array}{ccc} \rho_{1}  \\ \rho_{2}\\
\varphi_{1} \\
\end{array} \right)_{x},
\end{eqnarray}
where
\begin{align*}
B_{11}& = 3 \left(\rho_{1}^{2} - \varphi_{1}^{2} \right) -
\frac{3}{2} \rho_{1},\quad{} B_{12} = 0,\quad{}B_{13} = - 3
\rho_{1}
\varphi_{1}, \\
B_{21}& = \rho_{2} \left(6 \rho_{1} -3 \right) - 
\frac{9 \rho_{1}}{\rho_{2}} \varphi_{1}^{2},\quad{}B_{22} = 3 \left(\rho_{1} \left(\rho_{1}-1
\right) + 2 \rho_{2}^{2} - \varphi_{1}^{2}
 \right),\\
B_{23}& = - 6 \rho_{2} \varphi_{1},\quad{}B_{31} = \frac{3}{2}
 \varphi_{1} \left(4 \rho_{1} -1 \right),\quad{}B_{32} = \frac{3}{2}
 \frac{\rho_{1}^{2}}{\rho_{2}} \varphi_{1}  \left(\rho_{2} +1 \right),\\
B_{33}& = 3 \left(\rho_{1}^{2} - \varphi_{1}^{2} \right)-
\frac{3}{2} \rho_{1}.
\end{align*}

\subsection{Case III}
Let us note that symmetry constraint~(\ref{case3}) implies that
function $\phi$ must be real-valued, and we denote $
\left(\sigma_{1},\sigma_{2} \right) 
:= \nabla\phi$. Hence, the symmetry
constraint~(\ref{case3}) looks like
\begin{equation}
\label{case_III} u_{x} = \frac{1}{4} \left(\sigma_{1x} +
\sigma_{2y} \right).
\end{equation}
Moreover, for sake of simplicity, we assume function $S$
to be real-valued as well, and denote $ 
\left(\rho_{1},\rho_{2} \right):=\nabla S$. Differentiating
equation~(\ref{dVN_Lax_cartesian1}) with respect to $\lambda$, we
obtain the algebraic relation
\begin{equation}
\rho_{1} \sigma_{1} + \rho_{2} \sigma_{2} = 0,
\end{equation}
which allows us to eliminate, for instance, $\rho_{2}$. Using
these assumptions, we obtain the following hydrodynamic type system
in the variables $x$ and $y$, for the functions
$\sigma_{1}$, $\sigma_{2}$ and $\rho_{1}$
\begin{eqnarray}
\label{hydro_system4}
\left( \begin{array}{ccc} \sigma_{1}\\ \sigma_{2} \\ \rho_{1}    \\
\end{array} \right)_{y} =\left( \begin{array}{ccc}
0 & 1 & 0 \\ c_{1} & c_{2} & c_{3} \\
d_{1} & d_{2} & d_{3} \\
\end{array} \right)\left( \begin{array}{ccc} \sigma_{1}  \\ \sigma_{2}\\
\rho_{1} \\
\end{array} \right)_{x}
\end{eqnarray}
where
\begin{eqnarray}
&&c_{1}= 2 \frac{\sigma_{1}\rho_{1}^{2}}{\sigma_{2}^{2}}-1,
\quad{}c_{2}= - 2
\frac{\sigma_{1}^{2}\rho_{1}^{2}}{\sigma_{2}^{3}},
\quad{}c_{3}= 2 \rho_{1} \left(1+ \frac{\sigma_{1}^{2}}
{\sigma_{2}^{2}} \right), \nonumber\\
&&d_{1}= - \frac{\rho_{1}}{\sigma_{2}}, \quad{}d_{2}=
\frac{\sigma_{1}}{\sigma_{2}^{2}} \rho_{1},
\quad{}d_{3}=-\frac{\sigma_{1}}{\sigma_{2}}.
\end{eqnarray}
Using equation~(\ref{dVN_Lax_cartesian2}), the corresponding
equation for $\phi$, obtained by differentiation
of~(\ref{dVN_Lax_cartesian2}) with respect to $\lambda$ and the
system~(\ref{hydro_system4}), one gets the following expressions of
$V_{1}$ and $V_{2}$
\begin{eqnarray}
 \label{case3_V} &&V_{1} = - \frac{3}{2}
\sigma_{1} + \frac{3}{4} \left( \rho_{1}^{2} +
  \rho_{2}^{2}\right), \nonumber  \\
&&V_{2} = \frac{3}{2} \sigma_{2}.
\end{eqnarray}
Expressing $\rho_{2}$ in terms of $\sigma_{1}$, $\sigma_{2}$ and
$\rho_{1}$, one gets 
\begin{align}
\label{case3_Vbis}
V_{1}& = - \frac{3}{2} \sigma_{1} + \frac{3}{4} \rho_{1}^{2} +
\frac{3}{4} \frac{\rho_{1}^{2}\sigma_{1}^{2}}{\sigma_{2}^{2}} \\
V_{2}& =\frac{3}{2} \sigma_{2}.
\end{align}
Using the formula~(\ref{case3_Vbis}), one obtains
\begin{eqnarray}
\label{hydro_system5}
\left( \begin{array}{ccc} \sigma_{1}\\ \sigma_{2} \\ \rho_{1}    \\
\end{array} \right)_{t} =\left( \begin{array}{ccc}
C_{11} & C_{12} & C_{13} \\ C_{21} & C_{22} & C_{23} \\
C_{31} & C_{32} & C_{33} \\
\end{array} \right)\left( \begin{array}{ccc} \sigma_{1}  \\ \sigma_{2}\\
\rho_{1} \\
\end{array} \right)_{x}
\end{eqnarray}
where
\begin{align*}
C_{11}& = 3 \left(\frac{3}{2} \rho_{1}^{2} - \sigma_{1}
\right),\quad{}C_{12} = 3 \sigma_{2},\quad{}C_{13} = 9 \rho_{1}
\sigma_{1}, \\
C_{21}& = \frac{3\sigma_{1}^{2}}{\sigma_{2}^{2}} \rho_{1}^{4} \left (
\frac{1}{\sigma_{2}} -1\right) + \frac{3}{2} \sigma_{1} \rho_{1}^{2}
\left(1- \frac{3}{\sigma_{2}} \right) - 3 \sigma_{2}, \\
C_{22}& = \frac{3}{2} \rho_{1}^{2} \left(3 + 2
\frac{\sigma_{1}^{2}}{\sigma_{2}^{2}} \right) -
3\sigma_{1},\quad{}C_{23} = 3 \rho_{1} \sigma_{2} \left(2 + \frac{\sigma_{1}^{2}}{\sigma_{2}^{2}} \right) \\
C_{31}& = - 3 \rho_{1},\quad{}C_{32} = 0,\quad{}C_{33} = 3
\left(\rho_{1}^{2} - \sigma_{1} \right).
\end{align*}

Physical and geometrical meanings of the hydrodynamic type systems
obtained in this paper will be discussed elsewhere.\\

{\bf Acknowledgments.} L.\,V.\,B. is partially supported by RFBR grant 04-01-00508 and President of
Russia grant 1716-2003(scientific schools). B.\,G.\,K. and A.\,M. are supported in part by COFIN PRIN ``SINTESI'' 2002.


\begin{thebibliography}{99}

\bibitem{Zakharov} \paper{V.E. Zakharov}{Benney equations and
  quasi-classical approximation in the inverse problem
  method}{Funkts. Anal. Priloz.}{14}{89}{1980}.


\bibitem{Krichever1}
  \paper{I.M. Krichever}{Averaging method for
  two-dimensional integrable
  equations}{Func. Anal. Priloz.}{22}{37}{1988}.

\bibitem{Krichever2} \paper{I.M. Krichever}{The $\tau$-function of the
universal Whitham hierarchy, matrix models and topological field
theories}{Commun. Pure Appl. Math.}{47}{437}{1994}.


\bibitem{Dubrovin} \paper{B.A. Dubrovin and
  S.P. Novikov}{Hydrodynamics of weakly deformed soliton lattices:
  differential geometry and Hamiltonian theory}{Russian
  Math. Surveys}{44}{35}{1989}.


\bibitem{Singular} {\em Singular limits of dispersive waves},
  (eds. N.M. Ercolani et al.), Nato Adv. Sci. Inst. Ser. B Phys. {\bf
  320}, Plenum Press, New York (1994).

\bibitem{Ragnisco} \paper{B.G. Konopelchenko, L. Martinez Alonso and
  O. Ragnisco}{The $\bar{\partial}$-approach to the dispersionless KP
  hierarchy}{J. Phys. A: Math. Gen.}{34}{10209}{2001}.


\bibitem{Konopelchenko1} \paper{B. Konopelchenko and L. Martinez
  Alonso}{$\bar{\partial}$-equations,
integrable deformations of quasi-conformal mappings and Whitham
  hierarchy}{Phys. Lett. A}{286}{161}{2001}; 

\bibitem{Konopelchenko2} \paper{B.G. Konopelchenko
  and L. Martinez 
  Alonso}{Nonlinear dynamics on the plane and integrable hierarchies
  of infinitesimal deformations}{Stud. Appl. Math.}{109}{313-336}{2002}.

\bibitem{Kodama} \paper{Y. Kodama}{A method for solving the
  dispersionless KP equation and its exact
  solutions}{Phys. Lett. A}{129}{223}{1988}; Solutions of the
  dispersionless Toda equation, {\em Phys. Lett. A}, {\bf 147}, 477, (1990).

\bibitem{KodGibb} \paper{Y. Kodama and J. Gibbons}{A method for
  solving the dispersionless KP hierarchy and its exact
  solutions}{Phys. Lett. A}{135(3)}{167}{1989}.

\bibitem{Gibbons} \paper{J. Gibbons and S.P. Tsarev}{Conformal maps
  and reductions of the Benney equations}{Phys. Lett. A}{258}{263}{1999}.


\bibitem{Bogdanov} L.V. Bogdanov and B.G. Konopelchenko, {\em Symmetry
constraints for dispersionless integrable equations and systems of
hydrodynamic type}, {\tt arXiv:nlin.SI/0312013}, (2003).

\bibitem{ClassSymEng} \book{L.V. Ovsyannikov}{Group analysis of differential equations}
  {Nauka, Moscow}{1978}

%\bibitem{Olver} \book{P.J. Olver}{Applications of Lie groups to
%  differential equations}{Springer-Verlag, New York}{1993}.


\bibitem{Strampp} \paper{B. Konopelchenko, J. Sidorenko and
  W. Strampp}{1+1 dimensional integrable systems as symmetry
  constraints of 2+1 dimensional systems}{Phys. Lett. A}{157}{17}{1991}.

\bibitem{Cheng} \paper{Y. Cheng and
  Y.S. Li}{The constraint of the Kadomtsev-Petviashvili equation and
  its special solutions}{Phys. Lett. A}{157}{22}{1991}.

\bibitem{Segur} \book{M.J. Ablowitz and H. Segur}{Solitons and the
  Inverse Scattering Transform}{SIAM}{1981}.

\bibitem{Orlov} A. Yu. Orlov, Vertex operator,
  $\bar{\partial}$-problem, symmetries, variational identities and
  Hamiltonian formalism for 2+1 integrable systems, {\em Nonlinear and turbulent processes in
  physics}, ed. V. Baryakhtar, World Scientific, Singapore, 1988. 




\bibitem{Veselov} \paper{A.P. Veselov and S.P. Novikov}{Finite-zone
  two-dimensional potential Schr\"odinger operators. Explicit formulae
  and evolution equations}{DAN SSSR}{279}{20}{1984}.

\bibitem{Pinkall} \paper{B.G. Konopelchenko, U. Pinkall}{Integrable
  deformations of affine surfaces via Nizhnik-Veselov-Novikov
  equation}{Phys. Lett A}{245}{239-245}{1998}.

\bibitem{Ferapontov} \paper{E.V. Ferapontov}{Stationary
  Veselov-Novikov equation and isothermally asymptotic surfaces in
  projective differential geometry}{Diff. Geom. Appl.}{11}{117}{1999}.

\bibitem{KonopMoro} \paper{Boris G. Konopelchenko, Antonio
  Moro}{Geometrical optics in nonlinear media and integrable
  equations}{J.Phys. A: Math. Gen.}{37}{L105-L111}{2004}; Boris
  Konopelchenko and Antonio Moro, Integrable equations in
  nonlinear geometrical optics, to appear
  on Stud. Appl. Math., preprint {\tt arXiv:nlin.SI/0403051} (2004).




\end{thebibliography}
\end{document}